\documentclass[floatfix,amsmath,amsfonts,twocolumn]{revtex4}  
\usepackage{enumitem}
\usepackage{amsmath}
\usepackage{graphicx}
\usepackage{epsfig}
\usepackage{bm}
\usepackage{bbold}
\usepackage{amssymb}
\usepackage{color}

\newcommand{\bp}{{\bm p}}
\newcommand{\bP}{{\bm P}}
\newcommand{\bq}{{\bm q}}
\newcommand{\br}{{\bm r}}
\newcommand{\bs}{{\bm s}}

\newcommand{\pxx}{{\partial_x^2}}

\newcommand{\pz}{{\partial_z}}
\newcommand{\pzeta}{{\partial_\zeta}}

\newcommand{\pt}{{\partial_t}}

\newcommand{\ra}{\rangle}
\newcommand{\la}{\langle}

\newcommand{\z}{\zeta}
 
\newcommand{\rev}[1]{\textcolor{black}{#1}}

\begin{document}

\title{Transverse instability of hybrid solitons in the  strong light-matter coupling regime
}

\author{A. V. Yulin and  D. A. Zezyulin\footnote{email: d.zezyulin@gmail.com}}

\affiliation{School of Physics and Engineering, ITMO University, St. Petersburg 197101, Russia}

\date{\today}

\begin{abstract}
    We investigate the transverse instability of two-component  solitons forming in a planar \rev{waveguide} operating in the regime of  strong light-matter coupling.  The instability emerges as a result of the coupling  between transverse diffraction of the photonic component and  nonlinearity of the material excitations. Solutions   of three different forms are addressed which include  bright, gray-dark, and gray-gray solitons. In the limit of long-wavelength transverse perturbations, the instability is described with an  asymptotic expansion  whose predictions  agree with the results of  numerical simulations. The   dynamic development of instability of initially perturbed bright solitons leads to the formation of high-intensity spots in the photonic component. For gray-dark and dark-dark solitons, the transverse instability leads to the spontaneous nucleation of vortex-antivortex pairs which emerge in both fields as  transient patterns.   
\end{abstract}

\maketitle

\section{Introduction}

The regime of strong light-matter coupling occurs in the vicinity of the resonance between the frequency of the photons and that of the material excitations, provided that the losses are sufficiently small. In this regime the interaction results in the splitting between the real parts of the dispersion characteristics (in contrast to the the weak coupling regime, where the interaction results in the merging of the real parts of the dispersion characteristics while the imaginary parts are different). In semiconductor microcavities,  the hybridization between   photons and excitons enables the existence of exciton-polaritons \cite{review_QF_light,review_polaritons}. These quasiparticles have low effective mass and interact strongly, which results in the Bose-Einstein condensation of exciton-polaritons  at unusually high temperatures \cite{condensation_polaritons1,condensation_polaritons2,condensation_trap,condensation_room,Deng}. The discovery of this phenomenon has led to the development of a wide variety of polariton lasers \cite{review_polaritons}, including  lasers with electrical pump \cite{polariton_laser_e_pump}. 
 
Different localized nonlinear structures can form in the polariton lasers, including   vortices, half-vortices, Josephson vortices \cite{vortex_lattice,vortices_q,sol_vortices,half_vortices,v_half_v}. The exciton-polaritons can also be pumped directly by coherent light \cite{probing_supr_f,superfluid}. Dark and bright cavity solitons have been studied in these systems both theoretically and experimentally \cite{dark_s_res,bright_s_res,bright_s_res_2D,bright_s_res_exp}.  The polariton systems can serve as  highly nonlinear waveguides, where  femtosecond pulses  can transform into solitons with peak power of orders of magnitude  less than in silicon or GaAs waveguides \cite{prop_sol,prop_sol_Cher}.   
In most of the settings considered so far the characteristic energies of the nonlinear structures have been assumed to be much less compared to the width of the gap between the lower and the upper polariton dispersion branches. In this case the excitations belonging to the upper branch of the dispersion characteristic can be safely neglected, and the evolution of the field can be described by the slowly varying amplitude of the mode from the lower polariton branch. The resulting model is a well-studied generalization of the nonlinear Schr\"odinger equation, which in polariton physics is also known as the Gross-Pitaevski equation.  At the same time, in many experimental settings the nonlinear effects can be so strong that the upper polariton branch becomes populated as well.  Steady progress in the development of polariton platforms calls for a systematic investigation of polariton dynamics, and in particular, solitary waves, taking into account the contribution from both branches of the dispersion relation. 

Several previous studies have discussed the existence and stability of different types of solitons in systems that are described by the two branches resulting from the splitting of photon and exciton dispersions \cite{prop_sol_Cher,Karta2016,Yulin22,prop_sol,Egorov19,Walker}.  At the same time, most theoretical results have been obtained under the assumption that   the system is effectively one-dimensional (1D). However, many polariton platforms correspond to planar waveguides, where the dynamics in the second dimension cannot be ignored in general \cite{prop_sol,Walker,prop_sol_Cher}.  A famous example of a   situation where multiple dimensions lead to completely new physics is the transverse instability, which destroys a soliton localized in one (longitudinal) direction due to perturbations with wave vectors oriented in the second (transverse) direction. This fundamentally nonlinear effect  has been widely discussed  for  water waves, optical solitons,  plasma  solitons \cite{Kuzn1986,Peli2000,Kivshar}, and matter-wave solitons emerging in atomic Bose-Einstein condensates  \cite{Jones,KT88,Muryshev,Feder,Anderson,Brand,Kevrek2004,Kamchatnov}. Elongated dark solitons typically decay in arrays of isolated vortices or vortex rings \cite{Mamaev,Dutton,Anderson}, and hence the transverse instability may serve as   a  mechanism for the on-demand creation of vortices. However, regarding the transverse instability of polariton solitons, the results are much less abundant and are based on the Gross-Pitaevskii equation and its generalizations \cite{Smirnov,Zezyulin18,Zezyulin20}. 

In this paper, we investigate  the transverse instability of hybrid light-matter solitons that form in a planar microresonator operating in the regime of strong coupling. We consider a two-component model that incorporates the diffraction of the photonic component and the nonlinearity of material excitations in the second component. Hence the transverse instability arises specifically due to the strong coupling between light and excitons. For bright-bright and gray-dark solitons, we obtain analytical results for instability increment in the limit of  long-wavelength transverse perturbations.  Numerical simulations of nonlinear dynamics indicate that soliton stripes decay into soliton trains and arrays of vortices which exist as transient patterns. 

The rest of this paper is organized as follows. In Sec.~\ref{sec:model} we discuss the setup and the corresponding two-component model. In Sec.~\ref{sec:1d} we provide a brief overview of 1D solitons.  In Sec.~\ref{sec:equations} we derive the stability equations, and Sec.~\ref{sec:bright} and Sec.~\ref{sec:gray} present the main results for the transverse instability  of  bright-right and gray-dark solitons, respectively. Section~\ref{sec:concl} concludes the paper.

\section{The model}
\label{sec:model}

\begin{figure}
	\centering
	\includegraphics[width=0.999\columnwidth]{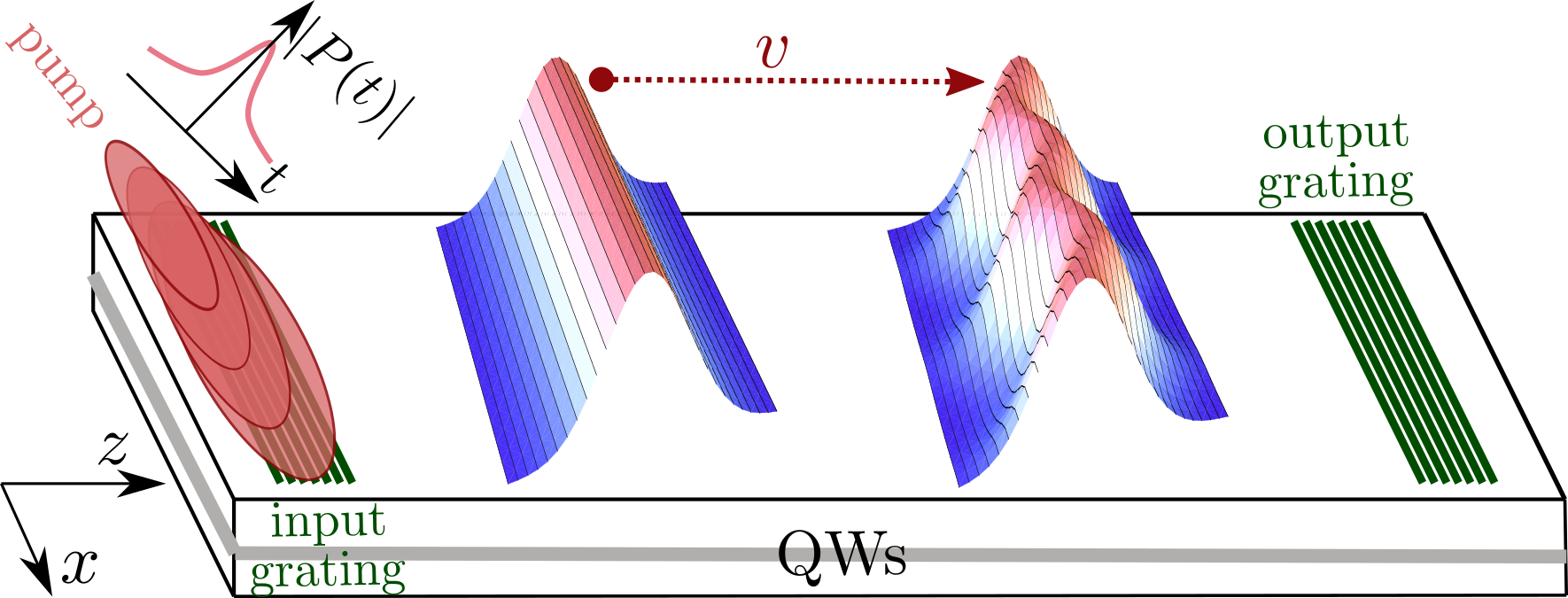}
	\caption{\rev{A schematic of the system. A planar waveguide with an embedded  layer of quantum wells (QWs)  is situated in the $(x,z)$ plane, where $x$ and $z$ correspond to  the transverse and longitudinal directions, respectively. The waveguide is pumped through the input grating with a pulse whose spatial aperture is wide in the transverse direction; the amplitude of the coherent pump is denoted as $|P(t)|$. The pulsed pump leads to the formation of an elongated soliton  whose shape  is  localized in the  $z$ direction and does  not depend on the transverse coordinate $x$. As the soliton propagates in the   longitudinal direction with velocity $v$, its shape  becomes modulated due to the transverse instability  resulting from the interplay between  nonlinearity and   diffraction along the $x$-axis. } }
	\label{fig:scheme}
\end{figure}   

\rev{In this paper, we consider a physical system shown schematically in  Fig.~\ref{fig:scheme}. The system consists of a thin semiconductor layer placed between two mirrors forming a planar waveguide for the photons. The existence of excitons is provided by a layer of quantum wells embedded in the waveguide. The thickness of the semiconductor layer is such that the frequency of the exciton resonance is higher than the cut-off frequency of the fundamental electromagnetic mode but lower then the cut-off frequencies of the higher-order modes. If the coupling strength $\Omega_R$ between the photons and the excitons exceeds the losses in the photon and exciton subsystems, then the regime of strong light-matter coupling is realized. Typically, the material dispersion of the semiconductor is small, and then dispersion of the electomagnetic mode can be approximated by the linear function, provided that $\Omega_R$ is much smaller compared to the detuning of the exciton resonance from the cut-off frequency of the photonic mode. } 

\rev{The system can be excited by a coherent pulse overlapped with CW radiation launched to the coupler that can be realized, for example, by periodic grating of the upper interface of the cavity, similar to as it is done in  \cite{prop_sol,{sol_prop_add1},{sol_prop_add2}}. The transverse width of the incident pulse's beam can significantly affect the formation and dynamics of polariton patterns. In particular, in \cite{prop_sol} the formation of the structures resembling spatial dark solitons has been observed.
The peculiarity of the present  paper is that we assume that the system is excited by a pulse whose  spatial aperture is  very wide in the transverse direction, and hence  the exciting field does not depend on the transverse coordinate.   }

In the strong coupling regime, photonic ($A$) and excitonic ($\psi$) fields are modelled by the  following two-component model \cite{prop_sol,Walker,Karta2016}:
\begin{equation}
	\label{eq:main}
i(\pt A + \pz A)  + \pxx A= -\kappa \psi, \quad i\pt \psi = -\kappa A + g |\psi|^2\psi.
\end{equation}
Here $t$ is the time normalized on some characteristic frequency $\Omega_{0}$, $z$ is the coordinate along the waveguide normalized on   $\Omega_0/v_g$, where $v_g$ is the group velocity of the pure photon mode at the resonant frequency of the material excitations $\Omega_m$.  In the equations (\ref{eq:main}) the coefficient $\kappa$ accounts for the light-matter coupling strength and  without loss of generality can be set to $1$. This means that the normalization frequency 
is chosen to be equal to   light-matter coupling rate. 

\rev{Let us remark that we use a relatively simple model, which, however, is able to provide  not only  a qualitatively but  also a quantitatively accurate description of polariton pulses propagating in GaAs waveguides  \cite{prop_sol}. To describe the evolution of polariton pulses in some other systems, for example in perovskite waveguides \cite{sol_prop_add2}, more complicated models are required to account for the effect of incoherent dark excitons and, possibly, of electron-hole plasma. The properties of the solitary waves in these systems can be quite different, and this case requires special consideration which is out of the scope of the present paper. Let us also note that the we focus on the propagation of the polaritons created by coherent optical pump in the system without incoherent pump leading to polariton condensation. 
}

\rev{The parameters of polaritonic waveguides can vary significantly. For GaAs waveguides \cite{prop_sol} the parameters can be estimated as follows: frequency of the exciton resonance $\omega_X=1.5$~eV, the Rabi frequency $\Omega_R=10$~meV, the group velocity $v_g=50$~$\mu$m/ps. By normalizing the time according to the Rabi frequency, we   obtain the normalization time $\tau_N \approx 65$~fs. The longitudinal coordinate   ($z$) is normalized to $L_z \approx 3$~$\mu$m and  the transverse   coordinate ($x$) is normalized  to $L_x \approx 0.2$~$\mu$m. The polariton lifetime can be as high as hundreds of picoseconds, and thus the conservative approximation considered in this paper is of physical importance. The flight time for the solitons in realistic cavities having length up to $1$~mm can be estimated as $>20$~ps. These numbers show that formation and evolution of the soltons discussed below can be observed in experiments with GaAs waveguides.   }

The function  $A(t,x,z)$ in (\ref{eq:main}) is the slow varying amplitude of the photon field and $\psi(t,x, z)$ is the order parameter   of the coherent excitons in the semiconductor microcavity.  We disregard the dispersion of the pure guided photons along the longitudinal direction $z$ assuming this is much smaller compared to the dispersion appearing due to the light-matter coupling (which agrees with experimental conditions \cite{prop_sol}).  The model also disregards photonic and excitonic losses, as we are   interested in the  dynamical development of  the instability at early stages rather by the long-term dissipative dynamics. The dissipation (and the pump) can be considered as a small perturbation  for wide class of the physically important exciton-polariton systems and then the conservative model gives a good approximation which can be useful for the understanding of the possible soliton solutions. These conservative solutions can also be used as a starting point for the perturbation theory accounting for the dissipative effects. 

The second derivative in  $x$ coordinate takes into account the diffraction   of the photonic field along the transverse direction.

 \section{Summary of 1D solutions}
\label{sec:1d}
 
\begin{figure}
	\centering
	\includegraphics[width=0.999\columnwidth]{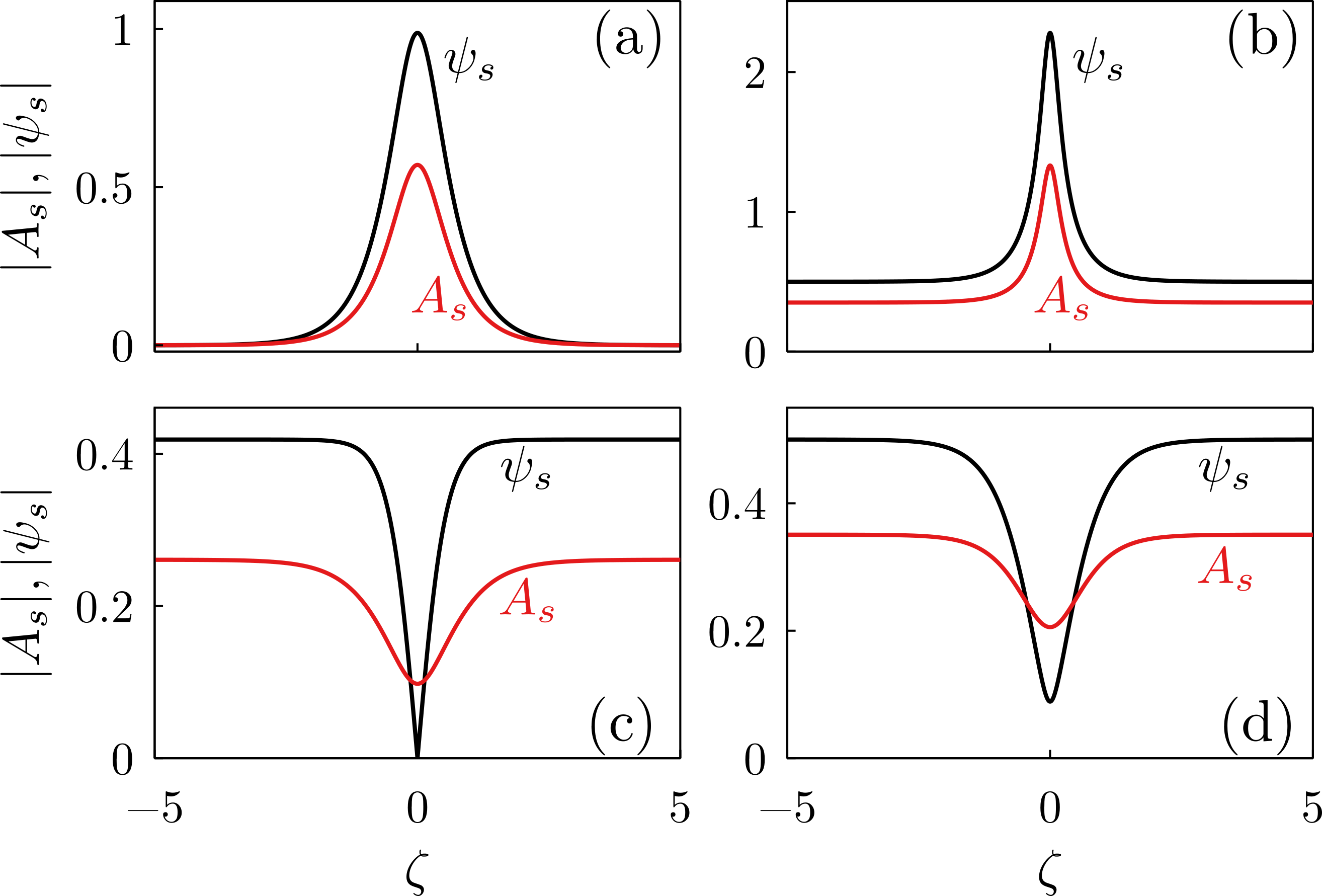}
	\caption{Examples of coexisting solutions of different types:    bright solitons on zero (a) and nonzero (b) backgrounds, gray-dark soliton  (c),  gray-gray soliton (d).  Black and red curves show moduli of photonic ($A$) and exitonic ($\psi$) fields. Horizontal axes $\zeta$ corresponds to   frames moving with velocity $v_s$. For all shown solutions, velocity corresponds to  $v_s=0.25$, and frequencies  are $\delta_s=-0.5$ (a), $\delta=1.07$ (b,d), $\delta=1$ (c). }
	\label{fig:examples}
\end{figure}

We start with a short summary of results pertaining to the  effectively 1D case,  when the diffraction in the transverse direction is absent, i.e., $\partial_x^2 A=0$. In this case, there exist solution steadily moving in the longitudinal $z$ direction. They can be found using the substitutions  $A  = A_s(\z) e^{-i\delta_s t}$,  and $\psi  = \psi_s(\z) e^{-i\delta_s t}$, where $\z =z - v_s t$  is the coordinate in the frame moving with velocity $v_s$, and $\delta_s$ is the frequency measured in the moving frame; subscript `$s$' stays for `soliton'. Fields $A_s$ and $\psi_s$ can be found from the system 
   \begin{eqnarray}
   \label{eq:stat}
   i(1-v_s)A_s' + \delta_s A_s = -\kappa \psi_s,\\
   \label{eq:stat2}
   -iv_s\psi_s' + \delta_s \psi_s = -\kappa A_s + g|\psi_s|^2\psi_s,
   \end{eqnarray}
  where primes denote derivative with respect to $\zeta$.   
   It has been found in \cite{prop_sol,Karta2016,Yulin22} that the  latter system has a variety of coexisting solitary-wave solutions which can be found in the analytical form. Representative examples of four solutions of different types are shown in Fig.~\ref{fig:examples}. It shows bright solitons on zero [Fig.~\ref{fig:examples}(a)]  and nonzero [Fig.~\ref{fig:examples}(b)] background, gray-dark soliton [Fig.~\ref{fig:examples}(c)]   and gray-gray soliton [Fig.~\ref{fig:examples}(d)]. The difference between  solutions of  two latter type consists in different properties of the excitonic component, whose amplitude becomes exactly zero for gray-dark solitons, but nowhere vanishes for gray-gray solitons. 
   
   In the earlier paper \cite{Yulin22} we have also studied stability of 1D solutions, which, in the context of the 2D model, corresponds to perturbations with the zero wavenumber and  infinite wavelength in the transverse $x$-direction. Bright solutions with nonzero background have been found unstable (see also \cite{Karta2016}), and  we disregard these solutions in what follows. Solutions of three other types have more complicated stability properties  and  can be stable or unstable, depending on the specific combination of velocity and frequency. In what follows, for illustration of our findings, we use solutions that are stable in the 1D geometry, and the eventual instability in the 2D geometry is caused specifically by a transverse perturbation of finite wavelength.

\section{Stability equations}
\label{sec:equations}

We consider perturbed stationary solutions in the form
\begin{eqnarray}
\label{eq:pert}
   A = e^{-i\delta_s t}[A_s(\z) +  a_1(\z)e^{ik_x x  + \lambda t} + a_2^*(\z)e^{-ik_x x  + \lambda^*t}],\\ 
   \psi = e^{-i\delta_s t}[\psi_s(\z) + \ p_1(\z)e^{ik_x x + \lambda t} + p_2^*(\z)e^{-ik_x x + \lambda^*t}],
\end{eqnarray}
where $a_{1,2}(\z)$ and $p_{1,2}(\z)$ describe the spatial shapes of the perturbations in the moving frame, and   and complex $\lambda$ characterizes   temporal behavior of the perturbation:  positive real part of $\lambda$ means that   the soliton is unstable. Real $k_x$ is the wavenumber of the transverse perturbation; the case $k_x=0$ recovers the stability equations for 1D solitons studied in \cite{Karta2016,Yulin22}.  Combining  substitutions (\ref{eq:pert}) with  Eqs.~(\ref{eq:main}) and keeping only linear (with respect to the small perturbations) terms, we arrive at the following spectral eigenvalue problem
\begin{equation}
    \label{eq:eig}
    (L_0 + k_x^2 L_2) \bp = i\lambda \bp,
\end{equation}
where $\bp = \bp(\z) = (a_1, a_2, p_1, p_2)^T$, $L_0$ is a $4\times 4$ block matrix  defined as  
\begin{eqnarray}
L_0 &=&  \left( \begin{array}{cc}
L_{11} -  \delta_s \sigma_3& -\kappa \sigma_3 \\
-\kappa \sigma_3 & L_{22} -  \delta_s \sigma_3 \end{array}\right),\\
L_{11} &=& \left( \begin{array}{cc}
 -i(1-v_s)  \pzeta  & 0\\[1.5mm]
 0 &-i(1-v_s)  \pzeta
 \end{array}\right),\\
L_{22} &=& \left( \begin{array}{cc}
 iv_s   \pzeta + 2g|\psi_s|^2 &  g\psi_s^2\\[1mm]
 - g(\psi_s^*)^2 &  iv_s   \pzeta  - 2g|\psi_s|^2 
 \end{array}\right),
\end{eqnarray}
where, for compactness, we use  $\sigma_3$ to denote the Pauli matrix. In addition, Eq.~(\ref{eq:eig}) includes the operator $L_2 = \textrm{diag}(1, -1, 0, 0)$ which takes into account the transverse perturbation.
%

\section{Bright solitons}
\label{sec:bright}

Bright solitons can be written down  in the compact form using   two auxiliary angles: $\alpha \in (0, \pi/2)$ and $\theta \in (-\pi/2, \pi/2)$, and adopting  the following parametrization for velocity and frequency:
\begin{equation}
\label{eq:parametrization1}
v_s = \sin^2\alpha, \quad \delta_s =  \kappa\sin(2\alpha)\sin \theta.
 \end{equation}
Then the following solution can be found  \cite{prop_sol,Yulin22}: 
\begin{equation}
\label{eq:bright}
\begin{array}{rcl}
\psi_s(\z) & =& \rho_s(\z) e^{-2i\z\kappa\cot(2\alpha)  \sin \theta   + i \Theta_s(\z) },\\[2mm]
A_s(\z) &=& \tan (\alpha)  \rho_s(\z)  e^{-2i\z\kappa\cot(2\alpha)  \sin \theta   + i\Theta_s(\z)/3 },
\end{array}
\end{equation}
where
\begin{equation}
\begin{array}{c}
\rho^2_s(\z) = \displaystyle \frac{4\kappa }{g} \frac{\tan\alpha\cos^2\theta}{  \cosh(4\kappa\cos\theta\csc(2\alpha)\z)-\sin \theta},\\[6mm]
\Theta_s(\z) = 
\displaystyle 3\arctan\left(\frac{1+\sin\theta}{\cos\theta} \tanh(2\kappa\cos\theta\csc(2\alpha)\z) \right).
\end{array}
\label{eq:bright2}
\end{equation}

\subsection{Transverse instability in the long-wavelength limit}

For  $k_x\to 0 $, one can develop the perturbation theory for weak instability with respect to  long-wavelength perturbations.   To construct the asymptotic expansion for unstable eigenvalues bifurcating from the origin,  we first summarize relevant information regarding the operator $L_0$. Its spectrum contains a zero eigenvalue $\lambda=0$ which is associated with  two linearly independent eigenvectors $\bp_{1,2}$: 
\begin{eqnarray}
\bp_1 &=& (A_s,\  -A_s^*,\ \psi_s,\  -\psi_s^*)^T,\\[2mm]
\bp_2 &=& (A_s',\   A_s'^{*},\  \psi_s',\  \psi_s'^{*})^T,
\end{eqnarray}
where we use superscript `$T$' to denote transposition. There also exist a pair of generalized eigenvectors  $\bq_{1,2}$ such that $L_0 \bq_{1,2} = \bp_{1,2}$. They  can be expressed   in terms of   derivatives of the   soliton   with respect to its frequency and velocity: 
\begin{eqnarray}
\bq_1 &=& \left(\frac{\partial A_s}{\partial \delta_s},\ \frac{\partial A_s^*}{\partial \delta_s},\ \frac{\partial \psi_s}{\partial \delta_s},\ \frac{\partial \psi^*_s}{\partial \delta_s}\right)^T,\\[2mm]
\bq_2 &=& i\left(\frac{\partial A_s}{\partial v_s},\ \frac{\partial A_s^*}{\partial v_s},\ \frac{\partial \psi_s}{\partial v_s},\ \frac{\partial \psi^*_s}{\partial v_s}\right)^T.
\end{eqnarray}
For two vectors $\bs_1 = \bs_1(\z)$ and $\bs_2 = \bs_2(\z)$ we introduce the inner product $\la \bs_1, \bs_2 \ra = \int_{-\infty}^\infty \bs_1^\dag(\z) \bs_2(\z)\, d\z$. Then one can define the    Hermitian adjoint   operator $L_0^\dag$. In its spectrum,  the zero eigenvalue $\lambda=0$ is associated with  two linearly independent eigenvectors
\begin{eqnarray}
\br_1 &=& (A_s,\  A_s^*,\ \psi_s,\  \psi_s^*)^T,\\[2mm]
\br_2 &=& (A_s',\   -A_s'^{*},\  \psi_s',\  -\psi_s'^{*})^T.
\end{eqnarray}
Moreover, eigenvectors from the kernels of $L_0$ and $L_0^\dag$ are mutually orthogonal, i.e.,
\begin{equation}
\label{eq:ortho}
\la \bp_i, \br_j \ra = 0 \quad  \mbox{ for any $i=1,2$ and $j=1,2$}. 
\end{equation}

Under the effect of the perturbation caused by small but nonzero $k_x$, the multiple  zero eigenvalue $\lambda=0$ in the spectrum of $L_0$ generically splits into several simple eigenvalues. If at least one of these eigenvalues has positive real part, then the soliton is unstable with respect to  the long-wavelength perturbations. We describe the behavior of the zero eigenvalue using  the expansion
\begin{equation}
\lambda = k_x \lambda_1 + k_x^2 \lambda_2 + O(k_x^3), \quad k_x\to 0,
\end{equation}
where $k_x$ is considered as a small parameter and  $\lambda_{1,2}$ are    coefficients to be determined. The corresponding eigenvector is sought in the form of the expansion
\begin{equation}
\bp =  \bP_0 + k_x \bP_1 + k_x^2 \bP_2 + O(k_x^3),
\end{equation}
where   $\bP_{0,1,2}$ are unknown vectors. 

Substituting the introduced expansions in the eigenvalue problem (\ref{eq:eig}) and collecting the terms at equal powers of  $k_x$, in the leading order  we obtain $L_0 \bP_0 = 0$, which implies  that $\bP_0$ should be sought as a linear combination of the eigenvectors from the kernel of $L_0$, i.e.,  
\begin{equation}
\label{eq:P0}
\bP_0 = \beta_1 \bp_1 + \beta_2 \bp_2,
\end{equation}
where   $\beta_1$ and $\beta_2$, $|\beta_1|^2+|\beta_2|^2\ne 0$, are unknown numerical coefficients.  In the next order, i.e., at  $k_x$, we obtain $L_0 \bP_1 = i\lambda_1 \bP_0$. The solution to the latter equation can be easily found as
\begin{equation}
\label{eq:P1}
\bP_1 = i\lambda_1(\beta_1\bq_1 + \beta_2\bq_2).
\end{equation}
The next order, $k_x^2$, gives the following equation
\begin{equation}
L_0 \bP_2 = i\lambda_2 \bP_0  + i\lambda_1 \bP_1 - L_2 \bP_0.
\end{equation}
This inhomogeneous equation has a solution only if its right-hand side is orthogonal to all eigenvectors from the kernel of the Hermitian adjoint operator $L_0^\dag$.   Multiplying (in the sense of the inner product $\la \cdot, \cdot \ra$) the right-hand side by $\br_1$ and $\br_2$, and using (\ref{eq:ortho}), (\ref{eq:P0}),  and (\ref{eq:P1}), we obtain a pair of homogeneous equations $B (\beta_1, \beta_2)^T=0$, where $B$ is a  $2\times 2$ matrix of the following form 
\begin{equation}
B=  \left(
\begin{array}{cc}
\lambda_1^2 \la \br_1, \bq_1 \ra + \la \br_1, L_2  \bp_1\ra      &   \lambda_1^2 \la \br_1, \bq_2 \ra + \la \br_1, L_2  \bp_2\ra \\[2mm]
\lambda_1^2 \la \br_2, \bq_1 \ra + \la \br_2, L_2  \bp_1\ra     &  \lambda_1^2 \la \br_2, \bq_2 \ra + \la \br_2, L_2  \bp_2\ra
\end{array}
\right).
\end{equation}
The solvability condition $\det B=0$ gives a biquadratic equation for $\lambda_1$ in the form $a \lambda_1^4 + b \lambda_1^2 + c = 0$, where $a$, $b$ and $c$ are the coefficients. Using the analytical  solution (\ref{eq:bright})--(\ref{eq:bright2}),  we compute 
\begin{eqnarray}
\label{eq:a}
a &=& (\pi+2\theta)^2 - 2(\pi+2\theta)\tan\theta - 3,\\[2mm]
b &=&  -{4\kappa \sin^3\alpha \cos\alpha \sin \theta}[(\pi+2\theta)\tan \theta + 2],\\[2mm]
c &=&  -{4\kappa^2\sin^6\alpha \cos^2\alpha }[(\pi+2\theta)^2 - 4\cos^2\theta].
\end{eqnarray}
For all $\alpha$ and $\theta$ in the region of existence of bright solitons, the determinant $b^2-4ac$ is positive, and the coefficient $c$ is negative. The leading coefficient $a$ is positive for $\theta\in(-\pi/2, \theta_\star)$ and negative for $\theta\in(\theta_\star, \pi/2)$, where $\theta_\star \approx 1.189$. In what follows, we consider only the former case, because in the latter one all bright solitons are strongly unstable in the 1D geometry \cite{Yulin22}.  For $a>0$   the biquadratic equation  has two purely imaginary and two purely real roots of opposite signs. The  positive root corresponds to instability, while the purely imaginary eigenvalues correspond to a pair of (stable) internal modes.

\begin{figure}
	\begin{center}
		\includegraphics[width=0.999\columnwidth]{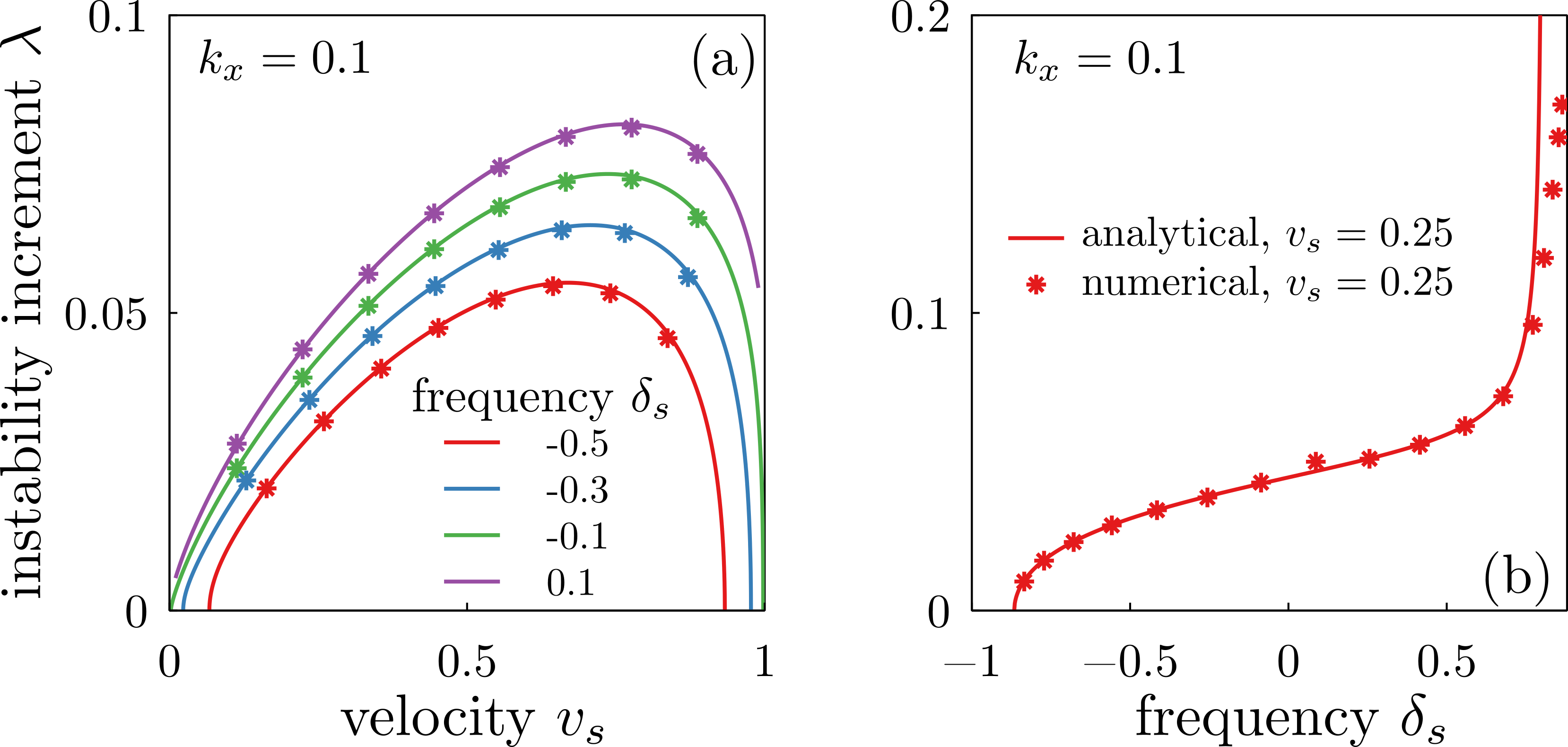}%
	\end{center}
	\caption{The increments of transverse instability obtained from the truncated asymptotic expansion (solid lines) and from the numerical solution of the spectral problem (asterisks) for fixed frequency $\delta_s$ and changing velocity $v_s$ (a) and for fixed velocity $v_s$ and changing frequency $\delta_s$ (b). In all cases the transverse perturbation wavenumber is set to $k_x=0.1$. }
	\label{fig:cmp}
\end{figure}  

To check the validity of the asymptotic result, we have computed the coefficient $\lambda_1$ for several representative combinations of parameters and compared     the instability increment obtained from the truncated expansion $\lambda \approx k_x \lambda_1$ with the analogous characteristic obtained from the  the direct numerical solution   of the eigenvalue problem (\ref{eq:eig}) at $k_x=0.1$. This comparison is illustrated in   Fig.~\ref{fig:cmp}  and shows good quantitative agreement between the analytical and numerical results. If the frequency $\delta_s$ is fixed, the instability rate is a nonmonotonous function of soliton velocity $v_s$.  On the other hand,  the   instability becomes stronger as the soliton frequency $\delta_s$ increases at fixed velocity $v_s$. Both these trends reflect that the instability increment  correlates with the   soliton amplitude which (as becomes evident from the analysis of exact solution) is a nonmonotonous function of velocity $v_s$ and grows with  frequency $\delta_s$.

\subsection{Transverse instability for finite wavelengths}
 
\begin{figure}
	\begin{center}
		\includegraphics[width=0.999\columnwidth]{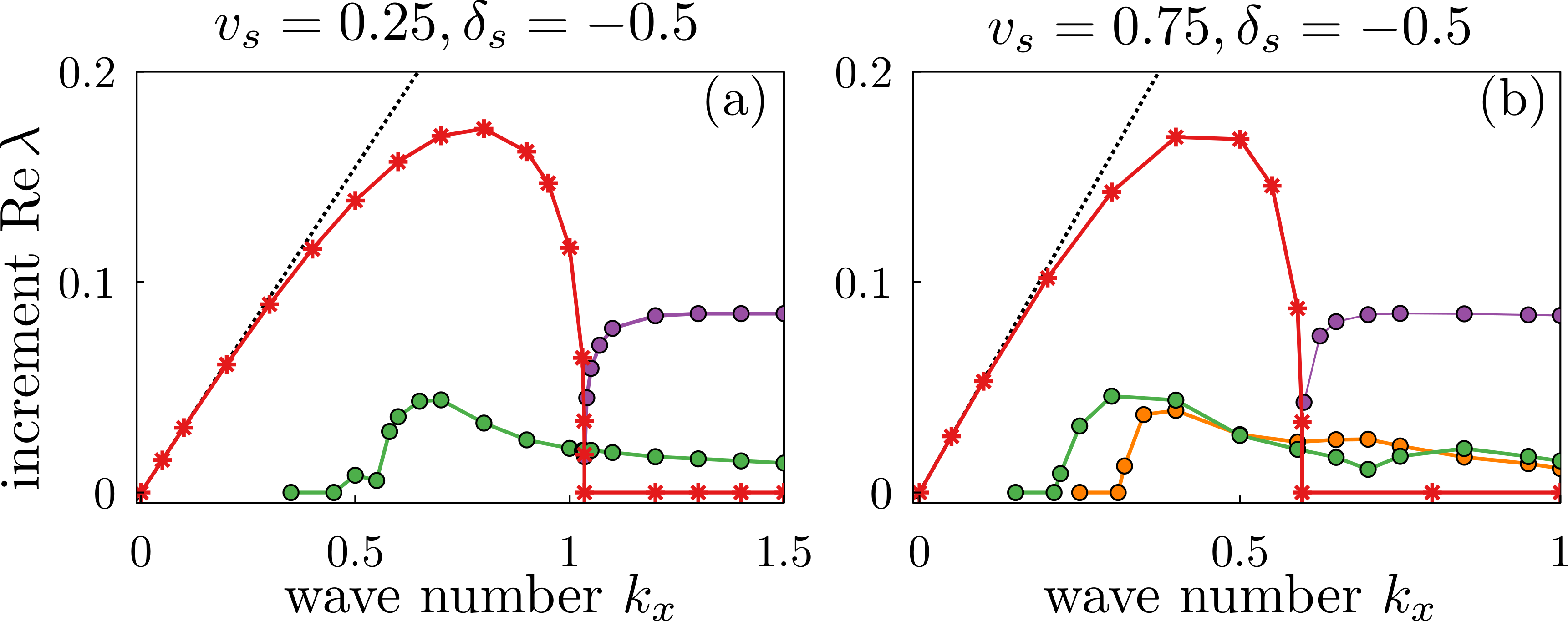}%
	\end{center}
	\caption{Instability increments for bright solitons plotted as functions of the perturbation wave number $k_x$.  	Dotted lines correspond to the linear dependence  predicted by the asymptotic expansions, and red asterisks correspond to the values obtained from the numerical soliton of the eigenvalue problem. Circles corresponds to additional instabilities which emerge with nonzero wavenumbers.  }
	\label{fig:kx}
\end{figure}

Next, we examine the behavior of the system under the increase of the wavenumber of the  perturbation   $k_x$. The corresponding plot of instability increment, obtained from the numerical solution of eigenvalue problem (\ref{eq:eig}) is presented in Fig.~\ref{fig:kx}. In agreement with the prediction of the asymptotic expansion, we observe that for small $k_x$ the instability increment   exhibits  the linear growth. However, for larger $k_x$ this   instability  becomes weaker and eventually disappears. The strongest instability is achieved at intermediate values of the wavenumber. For sufficiently large $k_x$, the system is additionally affected by secondary oscillatory instabilities (with complex eigenvalues $\lambda$) whose strength is smaller than the peak increment of the   instability predicted by the asymptotic expansion. 
 
\begin{figure}
	\centering
	\includegraphics[width=0.99\columnwidth]{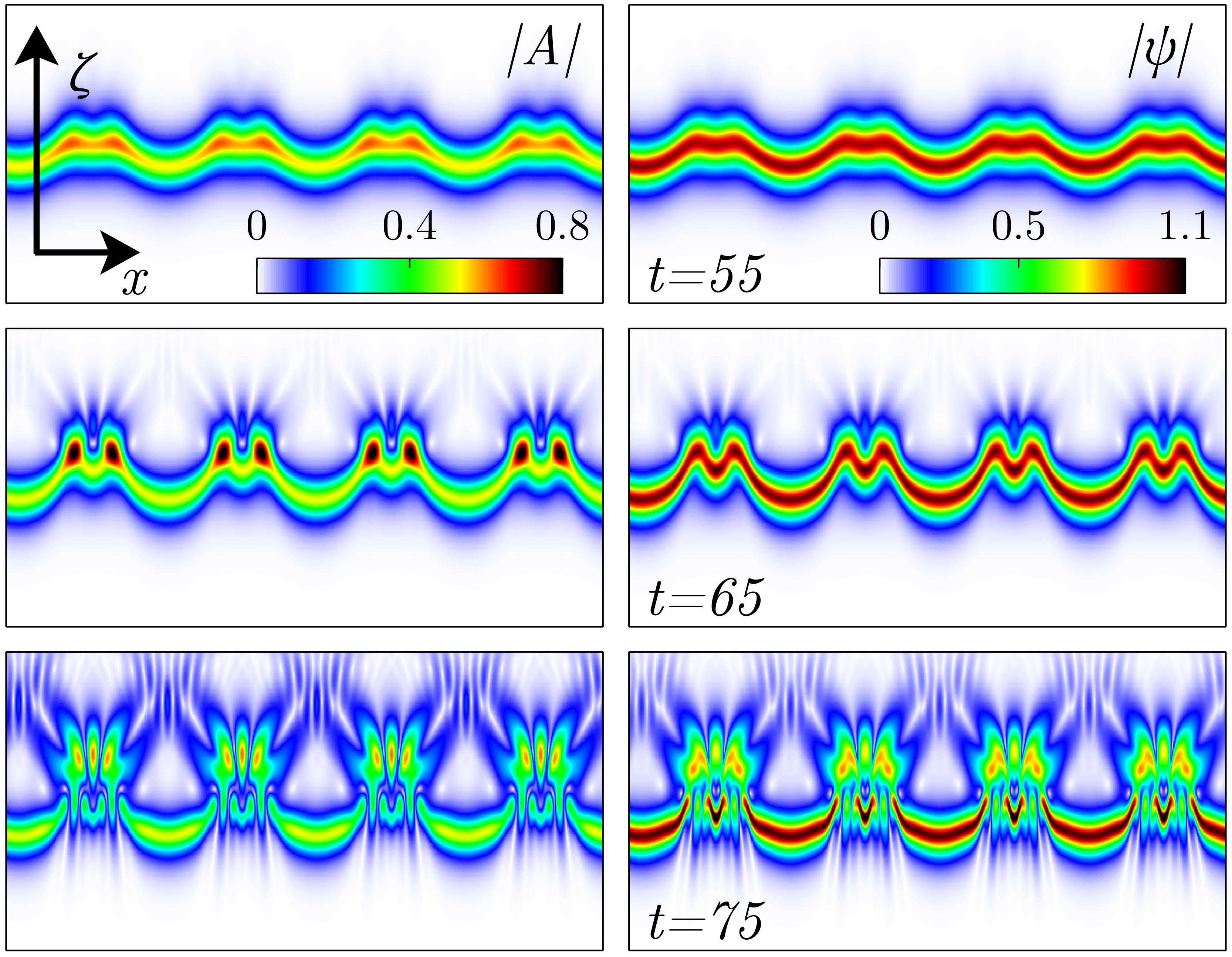}
	\caption{Decay of a bright-bright soliton with $v=0.25$ and $\delta=-0.5$. Left and right panels show moduli of photonic ($A$) and excitonic ($\psi$) fields.  The vertical $\zeta$-axis corresponds to the frame moving in the $z$-direction with velocity $v$. All plots are shown in the windows $(x, \zeta)\in [-63, 63]\times [-5.5,5.5]$. }
	\label{fig:bright01}
\end{figure}

\begin{figure}
	\centering
	\includegraphics[width=0.99\columnwidth]{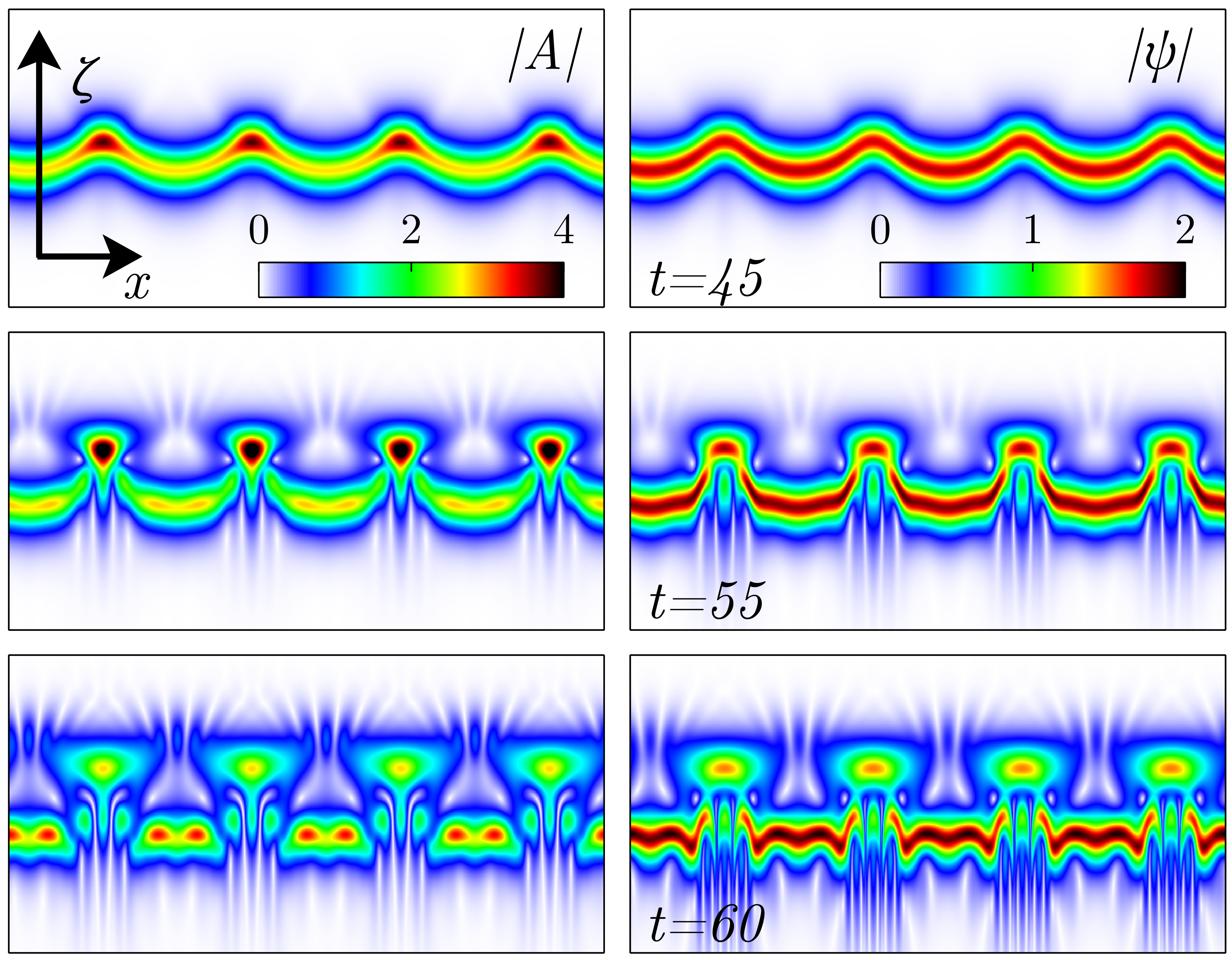}
	\caption{Decay of a bright-bright soliton with $v=0.75$ and $\delta=-0.5$. Left and right panels show moduli of photonic ($A$) and excitonic ($\psi$) fields.  The vertical $\zeta$-axis corresponds to the frame moving in the $z$-direction with velocity $v$. All plots are shown in the windows $(x, \zeta)\in [-63, 63]\times [-5.5,5.5]$.}
	\label{fig:bright02}
\end{figure}

Further, we have modelled the nonlinear stage of instability by solving system (\ref{eq:main}) with   a split-step pseudospectral method. Two representative examples of dynamically developing instability are presented in Fig.~\ref{fig:bright01} and \ref{fig:bright02}. They refer to initial conditions in the form of  two solitons with different velocities, namely $v_s = 0.25$ and $v_s=0.75$ coexisting at the same frequency $\delta_s=-0.5$. In either case the  instability has been introduced into the initial soliton  by a perturbation with $k_x =0.2$, and the width of the computation window in the transverse  $x$-direction was chosen to accommodate  four wavelengths of the perturbation.  At the early stages, the soliton stripe exhibits snaking which is accompanied by necking (i.e., the nucleation of several localized elevations of intensity in the photonic component). The number of high-amplitude spots can be different within the same spatial window (compare the panel with $t=65$ in  Fig.~\ref{fig:bright01} and with $t=55$ in Fig.~\ref{fig:bright02}).

\section{Gray-dark and gray-gray solitons}
\label{sec:gray}

 \begin{figure}[t]
	\begin{center}
		\includegraphics[width=0.999\columnwidth]{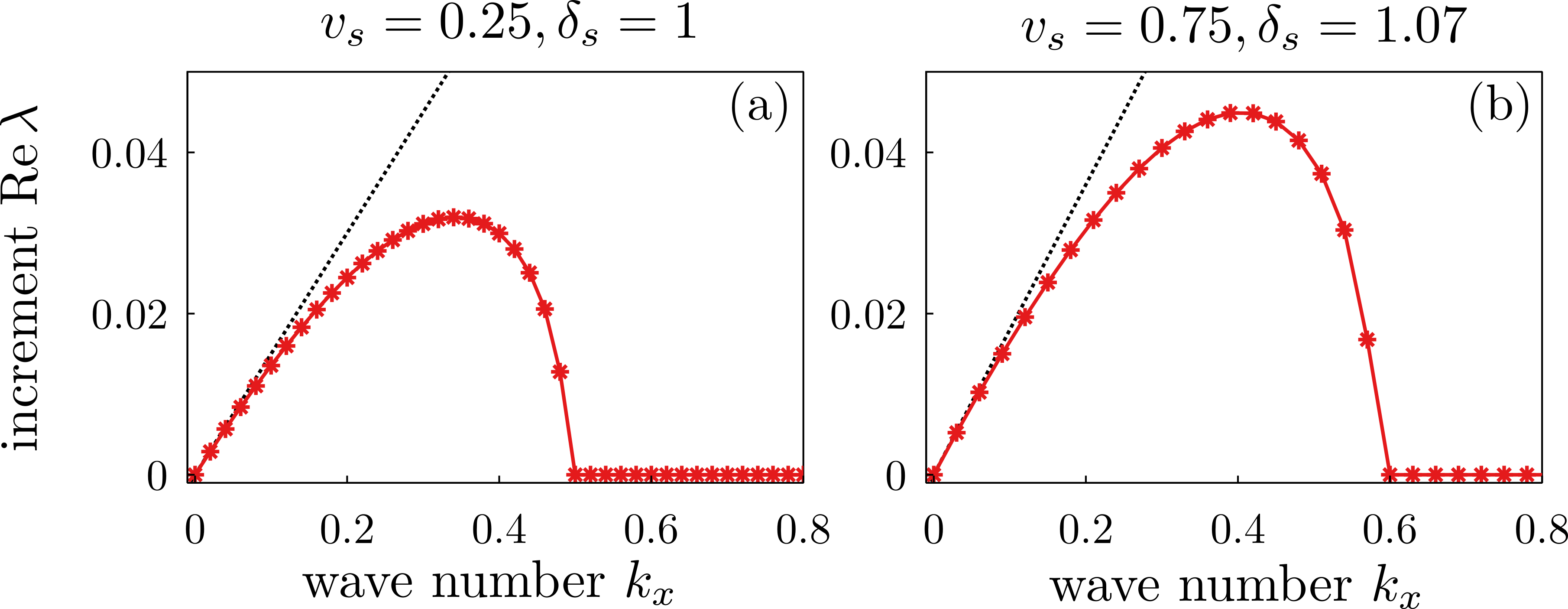}%
	\end{center}
	\caption{Instability increments for gray-dark (a) and gray-gray (b) solitons plotted as functions of the perturbation wavenumber $k_x$.  	Dotted line corresponds to the linear dependence  predicted by the asymptotic expansions, and red asterisks correspond to the values obtained from the numerical soliton of the eigenvalue problem.   }
	\label{fig:kxdark}
\end{figure}

To write down  dark-gray solitons solutions, for $v_s\in (0,1)$ it is convenient to   introduce the following parametrization for velocity and frequency:
\begin{eqnarray}
\label{eq:param2}
v_s =  \sin^2\alpha, \quad \delta_s = \frac{\kappa}{2}  \sin(2\alpha) (3e^\theta - e^{-\theta}),
\end{eqnarray}
where   $\alpha\in(0,\pi/2)$ and  $\theta>0$,
and the following constants
\begin{eqnarray}
\rho_\infty^2 = \frac{4 \kappa}{g} \tan\alpha \sinh \theta, \qquad  b = \frac{1}{4}(1-e^{-2\theta}), \nonumber \\ p=\kappa \csc(2\alpha)\sqrt{3e^{2\theta}-e^{-2\theta}-2}.
\end{eqnarray}
Then the kink-like  profile in the excitonic component reads
\begin{equation}
\label{eq:psidark}
\psi_s(\z) = \rho_s(\z)e^{-i\kappa(3e^{\theta} - e^{-\theta}) \cot(2\alpha)  \zeta+ i\Theta_s(\z)},
\end{equation}
where 
\begin{eqnarray}
\rho_s(\z) = \frac{\rho_\infty \sinh(p\z)}{\sqrt{\cosh^2(p\z) - b}}, \label{eq:dark}\\[3mm]
\Theta_s(\z) = 6\kappa\sinh\theta\csc(2\alpha) \z \nonumber\\- 3\arctan\left[\sqrt{\frac{1-e^{-2\theta}}{3+e^{-2\theta}}}\tanh(p\z)\right].
\label{eq:thetadark}
\end{eqnarray} 
The photonic component, $A_s$, can be determined from Eq.~(\ref{eq:stat2}).

For these solutions neither the fields $A_s$, $\psi_s$ nor their derivatives $A'_s$, $\psi'_s$ are localized. Using the exact solution, it easy to approximate the behavior of both field for large values of the moving-frame coordinate $\zeta$:
\begin{equation}
\{\psi_s, A_s\} = [\{\pm\rho_\infty, A_\infty\} + o(1)] e^{i[ \Omega \zeta \pm \Omega_0]},  \mbox{ as $\zeta\to \pm \infty$},
\end{equation}
where 
\begin{eqnarray}
\Omega =   \kappa\csc(2\alpha)\left[6\sinh\theta - (3e^\theta-e^{-\theta})\cos(2\alpha)\right],\\
\Omega_0 =    3\arctan\sqrt{\frac{1-e^{-2\theta}}{3+e^{-2\theta}}}.
\end{eqnarray}

The  eigenvectors $\bp_{1,2}$ and generalized eigenvectors $\bq_{1,2}$ introduced above in Sec.~\ref{sec:bright} can be understood only formally, because their entries do not represent square-integrable functions.  However, we  can construct linear combinations that correspond to  localized eigenfunction of operator $L_0$ and its adjoint $L_0^\dag$. Specifically, we set   $\bp = i\Omega \bp_1 - \bp_2$,  $\br = i\Omega \br_1 - \br_2$. In contrast to the case of bright solitons, in the case at hand the kernels of operators $L_0$ and $L_0^\dag$ are one-dimensional. Analysis of instability dark solitons in the nonlinear Schr\"odinger equation \cite{KT88} and also above analysis of instability bright solitons suggest  that the correction to the zero eigenvalue can be determined as
\begin{equation}
\label{eq:lambda2dark}
\lambda_1^2  = -\la \br, L_2  \bp \ra / \la \br, \bq \ra 
\end{equation}
where $\bq$ is a properly defined generalized eigenvector. It must not be localized, because the   formal scalar product $\la \br, \bq \ra $ is nevertheless  well-defined   due to the exponential decay of the entries of the   eigenvector  $\br$.    We compute the generalized vector $\bq$ numerically by solving linear equation $L_0 \bq = \bp$.  In Fig.~\ref{fig:kxdark} we  show that the results of this approach are in good agreement  with direct numerical evaluation of unstable eigenvalues for small $k_x$.

It is easy to see that  $\la \br, L_2  \bp \ra >0$, and hence the existence of instability is   determined by the sign of the denominator in (\ref{eq:lambda2dark}).  Computing $\lambda_1^2$ from  for several combinations of parameters, we observe that the denominator is negative. Hence $\lambda_1^2 >0$, and gray-dark solitons are unstable.    Similar analysis can also be conducted for gray-gray solitons, which also indicates the existence of instability. As follows from Fig.~\ref{fig:kxdark}, obtained from the numerical solution of the eigenvalue problem with different $k_x$, for both gray-dark and gray-gray solitons, the instability exists  in the bounded interval  of wavenumbers.

\begin{figure}
	\centering
	\includegraphics[width=0.999\columnwidth]{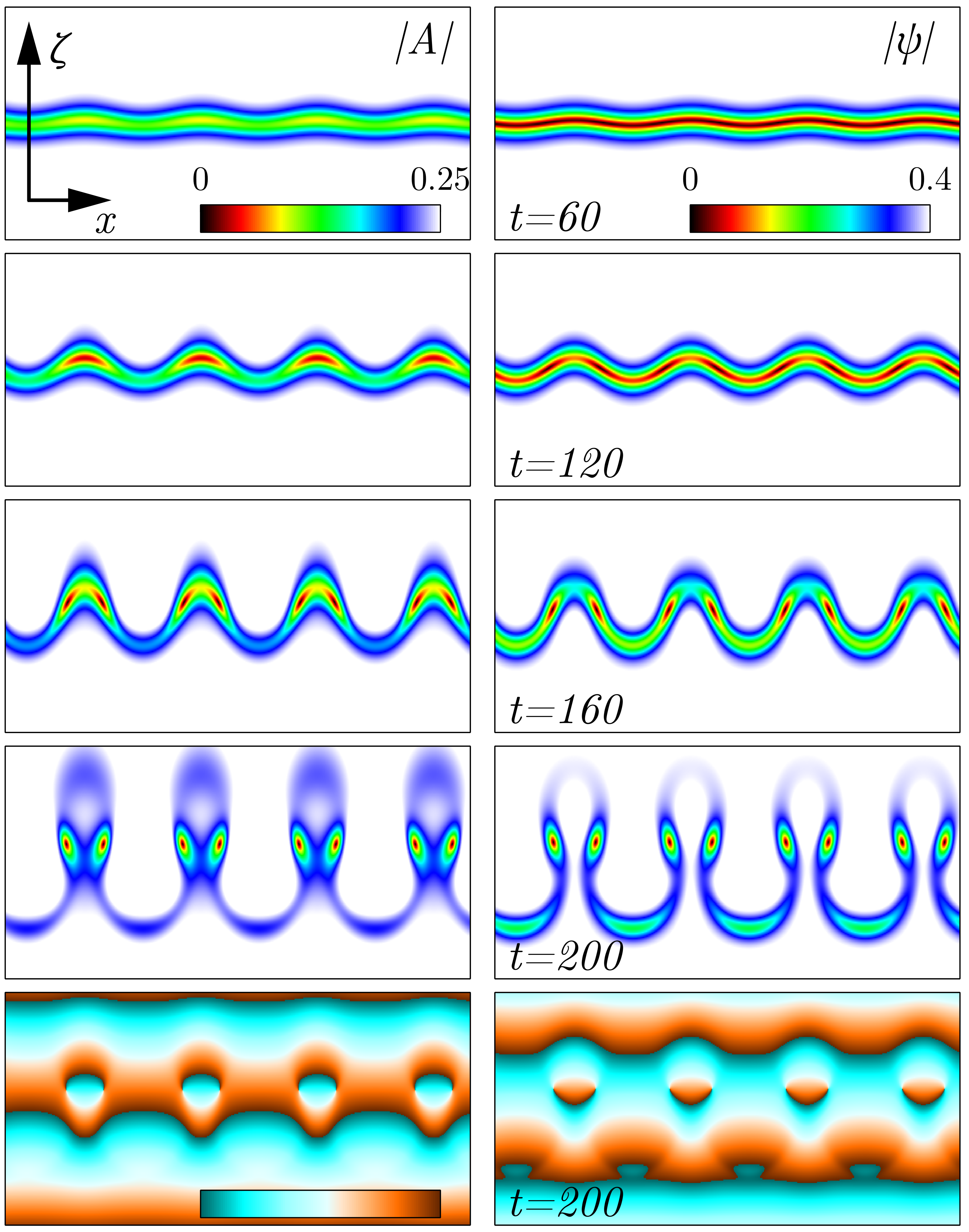}
	\caption{Decay of a gray-dark soliton with $v=0.25$ and $\delta=1$. Left and right panels show moduli of photonic ($A$) and excitonic ($\psi$) fields, except for two lower panels that show the phase distributions.  The vertical $\zeta$-axis corresponds to the frame moving in the $z$-direction with velocity $v$. All plots are shown in the windows $(x, \zeta)\in [-63, 63]\times [-8.6,8.6]$. }
	\label{fig:dark01}
\end{figure}
\begin{figure}
	\centering
	\includegraphics[width=0.99\columnwidth]{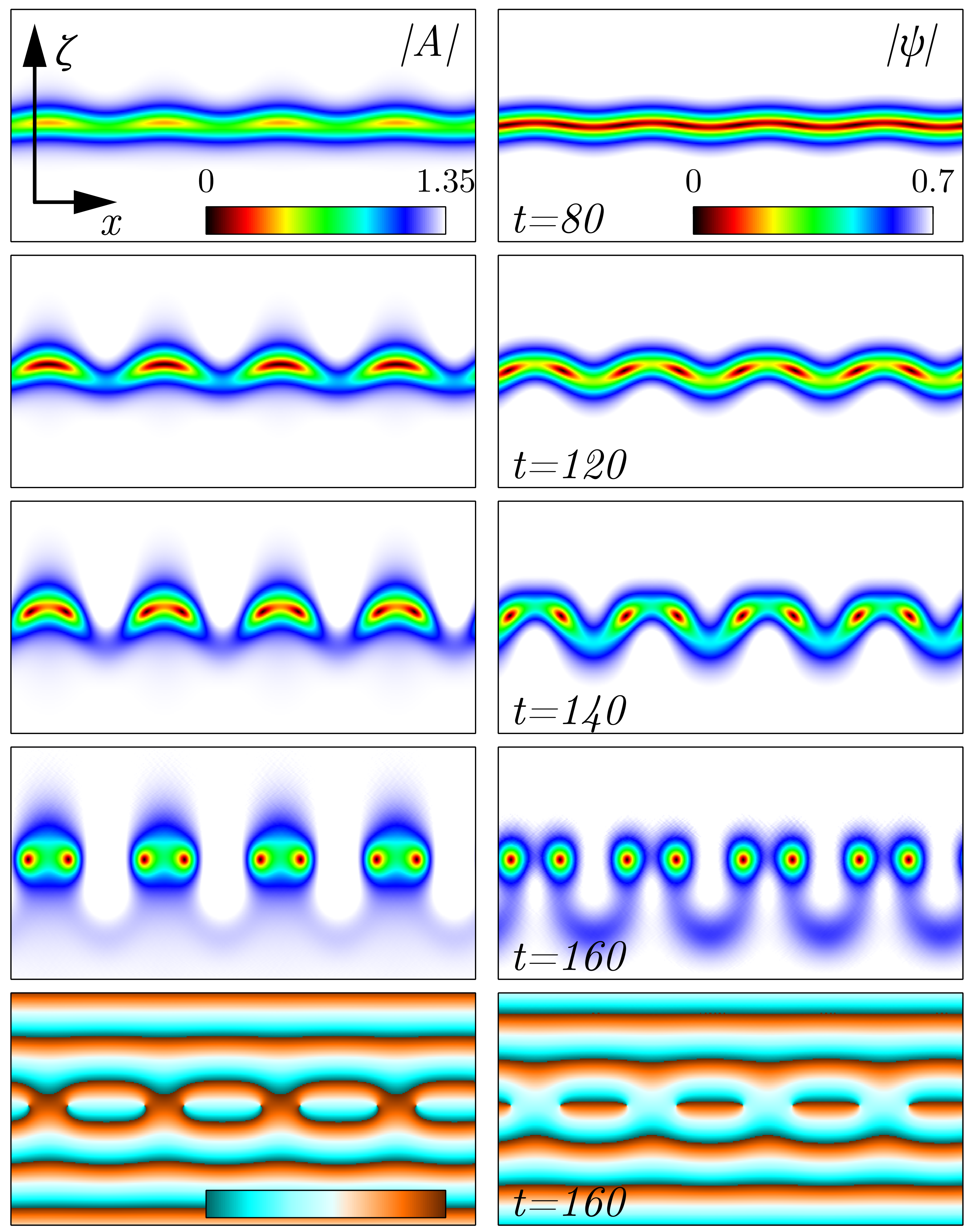}
	\caption{Decay of a gray-dark soliton with $v=0.75$ and $\delta=1.0$. Left and right panels show moduli of photonic ($A$) and excitonic ($\psi$) fields, except for two lower panels that show the phase distributions. The vertical $\zeta$-axis corresponds to the frame moving in the $z$-direction with velocity $v$. All plots are shown in the windows $(x, \zeta)\in [-63, 63]\times [-8.6,8.6]$. }
	\label{fig:dark02}
\end{figure}

Examples of dynamical simulations of slow and fast gray-dark solitons are shown in Fig.~\ref{fig:dark01} and \ref{fig:dark02}, respectively. The most representative feature which accompanies the dynamical bending of   dark-gray and gray-gray solitons is the spontaneous nucleation of vortices which occurs in both fields, despite  the intensity of the unperturbed gray photonic soliton nowhere vanishes.  Both components break  into   arrays of  vortex-antivortex pairs with topological charges of alternating signs. The  arrays of isolated vortices exist as transient patterns. For larger times, the neighbor vortices  tend to merge again, and the whole picture  becomes smeared by growing instability. Similar formation of vortex-antivortex pair also occurs for gray-gray solitons where both unperturbed fields do not have zeros.

\section{Conclusion}
\label{sec:concl}

In this paper, we have  studied the transverse instability of bright and solitons that form in a planar microcavity in the strong light-matter coupling regime. These solutions   have been modeled using a  two-component system which includes equations for excitonic and photonic fields and hence accounts for the excitations that belong to both branches of the polariton dispersion characteristic. The transverse instability emerges due to the combination of two factors that belong to different subsystems: the nonlinearity is associated with the excitonic components, and the transverse diffraction becomes relevant due to the present of the photonic component. 

Using the exact solutions for two-component bright and dark solitons, we have obtained analytical results for instability with respect to long-wavelength transverse perturbations. The persistence of instability for perturbations with finite wavelength has been verified numerically. We have also modeled the dynamic development of the instability. Perturbed bright solitons transform into transient patterns with localized elevations of the intensity, and dark solitons lead to formation of two-component vortices. \rev{These results naturally raise the question of the existence of two-component bright solitons and vortices that are localized in both spatial directions and stable on the timescale where the conservative approximation applies. In this context, we note that   patterns consisting of multiple bright peaks have been found to emerge in pumped polariton superfluids due to the development of the modulational instability \cite{Saito,Kwong}.} 
Another issue that deserves further investigation is the account of losses and corresponding dissipative dynamics that were not taken into consideration in this study.
 
\rev{The  theoretical model which have been adopted in the present study  takes into account two branches of the dispersion relation.    We finally note that a similar model can become  particularly relevant in the context of the dynamics of polaritons in periodically corrugated planar waveguides, where the gap opens due to Bragg scattering of polaritons.  These systems are now being actively studied, both theoretically and experimentally. This includes not just GaAs waveguides \cite{gap1,gap2}, but also perovskite platforms \cite{gap_perovskite}. In systems with periodic gratings, bound states in the continuum can form, which allows for the selection of a working mode. The gap width can then be controlled by the depth of modulation, and hence made quite narrow. 
The systems with small gap produced by the photon-exciton interaction can also be of interest.
}

\begin{acknowledgments}
The research is financially supported by Russian Science Foundation, Grant No. 25-12-00119 \cite{RNF}, and by Priority 2030 Federal Academic Leadership Program.

\end{acknowledgments}

\end{document}